\documentclass[a4paper,aps,onecolumn,nofootinbib]{revtex4}
\RequirePackage[colorlinks,hyperindex]{hyperref}
\RequirePackage[english]{babel}
\RequirePackage[latin1]{inputenc}
\RequirePackage[T1]{fontenc}
\RequirePackage{mathrsfs}
\RequirePackage{amsmath}
\RequirePackage{amssymb}
\RequirePackage{amsbsy}
\RequirePackage{color}
\RequirePackage{bm}
\hypersetup{colorlinks=true,breaklinks=true,urlcolor=blue,linkcolor=red}
\begin{document}
\title{\bf{Madelung Structure of the Dirac Equation}}
\author{Luca Fabbri$^{c}$\!\!\! $^{\hbar}$\!\!\! $^{G}$\footnote{luca.fabbri@unige.it}}
\affiliation{$^{c}$DIME, Universit\`{a} di Genova, Via all'Opera Pia 15, 16145 Genova, ITALY\\
$^{\hbar}$INFN, Sezione di Genova, Via Dodecaneso 33, 16146 Genova, ITALY\\
$^{G}$GNFM, Istituto Nazionale di Alta Matematica, P.le Aldo Moro 5, 00185 Roma, ITALY}
\date{\today}
\begin{abstract}
We consider the Dirac equations in polar form proving that they can equivalently be re-configured into a system of equations consisting of derivatives of the velocity density plus the Hamilton-Jacobi equation, giving the momentum in terms of relativistic quantum potentials (i.e. displaying first-order derivatives of the two degrees of freedom of the spinor field): this system is said to have Madelung structure. Conservation laws, second-order equations and multi-valuedness are also discussed.
\end{abstract}
\maketitle
\section{Introduction}
Of all interpretations of Quantum Mechanics, the Pilot-Wave Theory \cite{LdB}, or its more recent revisitation known as Bohmian Mechanics \cite{b1}, postulates that the wave function, whose behaviour is governed by the Schr\"{o}dinger equation, serves as the guide for the motion of a configuration of particles, encoded by the so-called guidance equation. Nowadays known as de Broglie-Bohm Theory, it is the only ontic interpretation of Quantum Mechanics involving hidden variables, as and such it is deterministic and non-local. The non-local hidden variables are the initial positions of the particles in configuration space. However, problems inherent to non-locality make it difficult to have this interpretation adapted to relativistic environments \cite{t1}. Even more so when a full treatment of relativistic spin is considered \cite{b2}. So far as we are aware the best attempts in this direction are those of Takabayasi in a series of references culminating in \cite{t2}.

The de Broglie-Bohm theory has at its basis the idea that the wave function be written in polar form\footnote{Here the term \emph{polar} form will always refer to the polar decomposition of complex functions into a module times an exponential of an imaginary angle, and never to the polar coordinates in the two-dimensional plane. Albeit the origin of the adjective 'polar' is the same for both cases, here we will never use any special coordinate system, working only in the most general curvilinear system of coordinates.} as the product of a module times a phase, respectively recognized as density and momentum, and that the Schr\"{o}dinger equation be correspondingly decomposed into hydrodynamic equations, that is one Hamilton-Jacobi equation and one continuity equation. When these two equations are taken together with the guidance equation, the whole system consists of the Newton equation and the conservation law for the mass, together known as Madelung equations \cite{Madelung:1927ksh}. The Madelung structure of quantum mechanics can consequently be reckoned as the attempt of converting the fundamental equations of non-relativistic quantum mechanics into a system of equations given by the conservation law of the mass and the Newton equation, both classical, and where the quantum characters are to be seen only in the fact that the Newton equation is sourced by a potential that contains two parts, one of which depending on the $\hbar$ constant. The extension to relativistic cases with spin was soon achieved by Yvon in \cite{Yvon1940}, although, surprisingly, in a formulation that was not manifestly covariant. This is perhaps the reason why following works of Takabayasi, collected in \cite{t2}, attracted more attention. In truth, however, also the works of Takabayasi, while impressive, fail to be the most complete of studies, as the treatment relies on rectilinear coordinates. Manifest general covariance for the polar form of relativistic spinor fields has been addressed in \cite{jl1, jl2}, two works cited by Takabayasi himself in a note added in proof in \cite{t2}. In this note, while clarifying that \cite{jl1, jl2} were written in parallel to \cite{t2}, he also admits that they might have helped him extend his own work, so that it is natural to expect Takabayasi to pick up the tools developed in \cite{jl1, jl2} to achieve the extension to curvilinear coordinates for the polar form of relativistic spinors. But as the Madelung re-formulation of quantum mechanics seems to approach its final steps, Takabayasi goes silent, and all advances along this path drop dead.

Whereas there is no doubt that the results of \cite{jl1, jl2} are one of the most important missing steps toward the realization of the full Madelung re-formulation of quantum mechanics, it is also certain that the results of \cite{jl1, jl2} alone cannot take the Madelung program to its end. In fact, while in \cite{jl1, jl2} Jakobi and Lochak give the most general expression for the polar form of relativistic spinor fields, they always keep their results algebraic and never take them to the differential level that one would need to implement the dynamics. To the best of our knowledge, this passage is done for the first time in reference \cite{Fabbri:2018crr}. Then, the possibility to write a general polar form for relativistic spinor fields as well as their covariant spinorial derivatives allows one to convert into polar form the Dirac relativistic spinor field equation, hence enabling the passage to the hydrodynamic form of the Dirac relativistic spinor field theory \cite{Fabbri:2023onb}.

Compared to the non-relativistic case, the relativistic case with spin has two differences, the first being that in a non-relativistic case we lack the definition of velocity that is instead naturally present in the relativistic case, so that the guidance equation that must be postulated in the former case is simply derived in the latter case \cite{Fabbri:2023yhl}.

The second difference is that in the non-relativistic spinless case the passage from the Hamilton-Jacobi equation to the Newton equation, done via a simple derivation, can always be reverted, by performing an integration, so that the Hamilton-Jacobi equation and the Newton equation are essentially equivalent. In the relativistic spinning case it does not make sense to talk about this equivalence because the passage from the Newton equation to the Hamilton-Jacobi equation cannot always be performed, since there exist no general ways to define integrals over spacetimes that may have non-trivial structures. Therefore, the most fundamental of the two is the Hamilton-Jacobi equation, and so as a consequence we will retain the right to define the Madelung equations as those containing a conservation law for the mass and a Hamilton-Jacobi equation. We will comment about the Newton law at the end \cite{Fabbri:2024avj}.
\section{Madelung Structure of the Schr{\"o}dinger Equation}\label{non-Rel}
We begin the treatment by recalling the generalities about non-relativistic spinless quantum mechanics, which will help us building the base-line that we will follow for the general case. In non-relativistic spinless quantum mechanics, the fundamental object is the wave function $\psi$ defined to be a complex scalar field. Any complex scalar can always be written in the form $\psi\!=\!\phi \exp{(iS)}$ where $\phi$ is a real scalar called module and $S$ is a real scalar called phase. Whereas the module squared $\psi^{*}\psi\!=\!\phi^{2}$ represents the density of the wave function, the phase can be interpreted recalling that, for plane-waves, which have constant density, we have $-i\vec{\nabla}\psi\!=\!\psi\vec{\nabla}S$, and that, for a quantum particle, the de Broglie condition reads $-i\vec{\nabla}\psi\!=\!\vec{P}\psi$ (the fact that the de Broglie condition define the momentum of the quantum particle is merely the expression of the relations $[x^{a},P_{b}]\!=\!i\delta^{a}_{b}$ with normalization $\hbar\!=\!1$): so $\vec{\nabla}S\!=\!\vec{P}$ in general. Such a condition and its temporal equivalent
\begin{gather}
\partial_{t}S\!=\!-H\label{act-t}\\
\vec{\nabla}S\!=\!\vec{P}\label{act-x}
\end{gather}
can be used to recognize in $S$ the action functional once we have that $H$ is the Hamiltonian and $\vec{P}$ is the momentum of the quantum particle (these conditions constitute the basis of quantum mechanics in its historical form, as discussed in the seminal works of de Broglie, Kennard and Heisenberg about the general structure of the wave function).

The dynamical features of non-relativistic spinless quantum mechanics are determined by the Schr\"{o}dinger equation
\begin{align}
i\partial_{t}\psi\!+\!\frac{1}{2m}\vec{\nabla}\!\cdot\!\vec{\nabla}\psi\!=\!V\psi\label{S}
\end{align}
in which $V$ a generic real scalar potential. In polar form $\psi\!=\!\phi \exp{(iS)}$ and using (\ref{act-t}-\ref{act-x}), it decomposes as
\begin{gather}
\partial_{t}(m\phi^{2})\!+\!\vec{\nabla}\!\cdot\!(\phi^{2}\vec{P})\!=\!0\label{contpolar}\\
H\!=\!\frac{1}{2m}\vec{P}\!\cdot\!\vec{P}
\!+\!\left(V\!-\!\frac{1}{2m}\phi^{-1}\vec{\nabla}\!\cdot\!\vec{\nabla}\phi\right)\label{enerpolar}
\end{gather}
which can respectively be identified with a continuity equation and a type of Hamilton-Jacobi equation with a potential given by the sum of an external potential $V$ plus the quantum potential $Q\!=\!-\vec{\nabla}\!\cdot\!\vec{\nabla}\phi/(2m\phi)$ \cite{b1}. It is left as an exercise for the reader to see that they imply back the Schr\"{o}dinger equation, and so they are equivalent to the Schr\"{o}dinger equation itself. Thus, the Schr\"{o}dinger equation can be converted into a continuity equation, which is classical, and a Hamilton-Jacobi equation, also classical in form, and it is only in the fact that the potential is shifted by the quantum potential $Q$ that quantum effects are found \cite{b1}. Such an occurrence lies at the basis of the philosophical current that aims at interpreting quantum mechanics in terms of classical pictures, as was originally described in \cite{t1} and \cite{b2, t2}.

On the other hand, while the continuity equation (\ref{contpolar}) has the structure of a conservation law, one may ask \emph{of what} this is a conservation law. The most natural thing would be the mass, but in order for (\ref{contpolar}) to be the mass conservation law $\partial_{t}(m\rho)\!+\!\vec{\nabla}\!\cdot\!(m\rho\vec{v})\!=\!0$ with $\rho$ the density and $\vec{v}$ the velocity, one must also insist on the identification
\begin{gather}
\phi^{2}\!=\!\rho\label{Born}\\
\vec{P}\!=\!m\vec{v}\label{momentum}:
\end{gather}
while the first can just be taken as the definition of the density, the second accounts for a constraint tying momentum and velocity which has to be assumed, no matter how natural it might look. And there are physical situations where (\ref{momentum}) is not correct, as we are going to see in the relativistic case. If (\ref{act-x}) is taken into (\ref{momentum}) we obtain
\begin{align}
\vec{\nabla}S\!=\!m\vec{v}\label{guidance}
\end{align}
called \emph{guidance equation}. That (\ref{momentum}) is assumed corresponds to the fact that the guidance equation (\ref{guidance}) must be assumed, no matter how natural it might look in view of the previous arguments. Again, it is not obvious that the velocity can be the gradient of some scalar potential, and in fact, one can find works where (\ref{guidance}) is criticized \cite{Wallstrom:1994fp} (at times, instead of saying that the velocity is the gradient of some scalar potential one can write the more intrinsic $\vec{\nabla}\!\times\!\vec{v}\!=\!0$ \cite{Reddiger:2015vsa}).

At this stage, the Schr\"{o}dinger equation (\ref{S}) plus the guidance equation (\ref{guidance}) together are equivalent to the system
\begin{gather}
\vec{P}\!=\!m\vec{v}\\
\partial_{t}(m\rho)\!+\!\vec{\nabla}\!\cdot\!(m\rho\vec{v})\!=\!0\\
H\!=\!\frac{1}{2m}\vec{P}\!\cdot\!\vec{P}\!+\!(V\!+\!Q)
\end{gather}
as the guidance equation, the conservation law of the mass and the Hamilton-Jacobi equation, respectively.

A final step is to take the gradient of the Hamilton-Jacobi equation getting
\begin{align}
\partial_{t}\vec{P}\!+\!\vec{v}\!\cdot\!\vec{\nabla}\vec{P}\!=\!-\vec{\nabla}(V\!+\!Q)
\end{align}
which is the Newton law. Because this can be integrated back to the Hamilton-Jacobi equation, we conclude that the Hamilton-Jacobi equation and the Newton law are equivalent. Hence, we can take the fundamental system as the one given by guidance equation, mass conservation law and Newton law, and this is what is said to be in Madelung form.

Before proceeding, we have to discuss two points, as anticipated in the introduction, about what we should expect in going from non-relativistic to relativistic cases. One point is that, as we just discussed, in the non-relativistic case one must assume the guidance equation (\ref{guidance}) to be true. The reason is that in the non-relativistic cases there exists no well-defined way to introduce the velocity. In relativistic case instead there is a definition of velocity, as we will see in the following. Consequently, one might expect that the guidance equation be derived within the theory. We shall see that this is the case. And therefore, this will also indirectly reply to the question that was raised in reference \cite{Hatifi:2024pks}.

The other point is that, again as discussed above, in the non-relativistic case the Hamilton-Jacobi equation can be obtained from the Newton equation via an integration, but in the relativistic case if the manifold is generally curved such an integration is not always well-defined and thus the Hamilton-Jacobi equation can not be obtained from the Newton equation. As we want to work in the most general of cases, and because in such a situation the Hamilton-Jacobi equation cannot be obtained from the Newton equation, although the converse is always true, the Hamilton-Jacobi equation is more fundamental than the Newton equation. Consequently, generality asks that the Madelung form be defined as the system containing a guidance equation, a mass conservation law and a Hamilton-Jacobi equation.

So, we can define the \emph{Madelung form} as the system of equations that contains a guidance equation, a conservation law of the mass and a Hamilton-Jacobi equation, \emph{i.e. all equations that are formally classical}. Our primary target is to find a system of equations in the Madelung form that is as a whole equivalent to the Dirac equation.

However, before doing that in the full $(1\!+\!3)$-dimensional spacetime, it may be instructive to see what happens in lower-dimensional spaces. Once accustomed with the basic ideas, it will be easier to generalize.
\section{(1+2)-dimensional spacetime}\label{1+2}
Because the above section treats the non-relativistic case, which is in a $3$-dimensional space, we will start from this dimensionality. However, because we want a geometry that resembles the one we are planning to study later, we shall not consider $3$-dimensional spaces, but $(1\!+\!2)$-dimensional spacetimes. In any dimension, the Clifford gamma matrices $\boldsymbol{\gamma}^{a}$ are used to define the sigma matrices as $\boldsymbol{\sigma}_{ik}\!=\![\boldsymbol{\gamma}_{i},\boldsymbol{\gamma}_{k}]/4$ although, in the present dimension and signature, we have that $2\boldsymbol{\sigma}^{ab}\!=\!i\varepsilon^{abc}\boldsymbol{\gamma}_{c}$ in general. Given a spinor, its adjoint is $\overline{\psi}\!=\!\psi^{\dagger}\boldsymbol{\gamma}^{0}$, and given the pair of adjoint spinors, we define the spinorial bi-linears $U^{a}$ as velocity density vector and $\Phi$ as density scalar following \cite{Fabbri:2024lyu}. When the spinor field is re-written in polar form, these can be re-expressed as $\Phi\!=\!\phi^{2}$ and $U^{a}\!=\!\phi^{2}u^{a}$ in terms of the module $\phi$ and the velocity $u^{a}$ verifying $u_{a}u^{a}\!=\!1$ as normalization. The module is the unique degree of freedom of the spinor field.

In reference \cite{Fabbri:2024lyu} it was shown that we can always define a tensor $R_{ij\mu}$ and a vector $P_{\mu}$, respectively called \emph{spacetime and gauge tensorial connections}, in terms of which we can extend the polar decomposition of spinor fields also to their covariant derivatives. In particular
\begin{eqnarray}
&\nabla_{\mu}u_{i}\!=\!R_{ji\mu}u^{j}\label{3du}
\end{eqnarray}
can be proven as a general identity.

In \cite{Fabbri:2024lyu} it was also discussed that the Dirac equation can be equivalently re-written as
\begin{eqnarray}
&\frac{1}{2}R_{ija}\varepsilon^{ija}\!+\!2P^{k}u_{k}\!-\!2m\!=\!0\label{constraint}\\
&R_{k}\!+\!2\varepsilon_{kab}P^{a}u^{b}\!+\!\nabla_{k}\ln{\phi^{2}}\!=\!0\label{trueequation}
\end{eqnarray}
in which we notice a peculiar occurrence. In three dimensions, for whatever signature, the Dirac equation consists of two complex differential equations, and so four real differential conditions, to be satisfied. On the other hand, in three dimensions, a spinor has a unique degree of freedom, which is fully determined, with its three derivatives, employing only the three Dirac equations (\ref{trueequation}). As for the remaining Dirac equation (\ref{constraint}), it has become a constraint \cite{Fabbri:2024lyu}.

Now, with this formalism, it is possible to re-arrange the Dirac equations in polar form (\ref{constraint}-\ref{trueequation}) according to
\begin{eqnarray}
&\nabla_{\mu}U^{\mu}\!=\!0\label{M3-1}\\
&P^{i}\!=\!(m\!-\!\frac{1}{4}R_{abc}\varepsilon^{abc})u^{i}
\!-\!\frac{1}{2}\varepsilon^{ijk}u_{j}(\nabla_{k}\ln{\phi^{2}}\!+\!R_{k})\label{M3-2}
\end{eqnarray}
as shown in appendix \ref{app1}. The first is the continuity equation for the velocity density. The second tells that what was called gauge tensorial connection is simply the momentum of the quantum particle. But more importantly, it tells us that the momentum can always be written explicitly in terms of the spacetime tensorial connection and a term of the type $\nabla_{k}\phi/\phi$ up to proportionality factors. Always up to irrelevant factors, there is a complete analogy between this term and the quantum potential $Q$ of the non-relativistic case: the only difference is that this term is of the first-order derivative while $Q$ is of the second-order derivative, and such a difference can be reduced to the Dirac equation being first-order derivative as opposed to the Schr\"{o}dinger equation being second-order derivative.

Therefore $\nabla_{k}\ln{\phi^{2}}$ is the relativistic expression of the quantum potential. Then, equation (\ref{M3-2}), tying the components of the momentum to the quantum potential, is by its very construction the relativistic form of the Hamilton-Jacobi equation. Equation (\ref{M3-2}), in tying the momentum to the product of mass and velocity, is also the relativistic expression of the guidance equation. The fact that Hamilton-Jacobi equation and guidance equation are the same equation comes again as a feature of relativistic situations since in relativistic cases the dispersion relations are linear.

As equation (\ref{M3-1}) is the continuity equation, and since the full system (\ref{M3-1}-\ref{M3-2}) is equivalent to (\ref{constraint}-\ref{trueequation}), the system of equations (\ref{M3-1}-\ref{M3-2}) is in Madelung form. The fact that the Madelung system is made of the continuity equation and the Hamilton-Jacobi/guidance equation may seem a rather general circumstance, at this stage. However, as we shall see, it is a distinctive feature of $3$-dimensional spaces only. Before, tackling our main target, that is the $(1\!+\!3)$-dimensional spacetime, we will see what happens in the $(1\!+\!1)$-dimensional spacetime, as there is a lesson to be learned.
\section{(1+1)-dimensional spacetime}\label{1+1}
In the $(1\!+\!1)$-dimensional spacetime the structure of the theory is, surprisingly, more complicated than the one seen in the previous section, due to the presence of parity-odd objects. Clifford gamma matrices and sigma matrices will be defined as usual, but in this dimension and signature, we have that $2\boldsymbol{\sigma}_{ab}\!=\!\varepsilon_{ab}\boldsymbol{\pi}$ defining the matrix $\boldsymbol{\pi}$ (the symbol $\boldsymbol{\pi}$ stands for the Greek letter $\pi$, that is $p$, since this matrix is parity-odd). The spinorial bi-linears are $U^{a}$ as velocity density vector, $\Theta$ as density pseudo-scalar and $\Phi$ as density scalar again following \cite{Fabbri:2024lyu}. In the polar form $\Theta\!=\!2\phi^{2}\sin{\beta}$ and $\Phi\!=\!2\phi^{2}\cos{\beta}$ with $U^{a}\!=\!2\phi^{2}u^{a}$ in terms of module $\phi$ and chiral angle $\beta$ and with the velocity $u^{a}$ verifying $u_{a}u^{a}\!=\!1$ as normalization condition. The module and chiral angle are the two degrees of freedom of the system.

The spacetime and gauge tensorial connections are defined as usual, and so the covariant derivative of spinor fields in polar form can also be given \cite{Fabbri:2024lyu}. And again as usual we have that
\begin{eqnarray}
&\nabla_{\mu}u_{i}\!=\!u^{j}R_{ji\mu}\label{2du}
\end{eqnarray}
is still a valid general identity although now the indices run over two values only.

The Dirac equation is equivalent to
\begin{eqnarray}
&\nabla_{\mu}\beta\!-\!2P^{\alpha}\varepsilon_{\alpha\mu}
\!+\!2mu^{\alpha}\varepsilon_{\alpha\mu}\cos{\beta}\!=\!0\label{chan}\\
&\nabla_{\mu}\ln{\phi^{2}}\!+\!R_{\mu}
\!+\!2mu^{\alpha}\varepsilon_{\alpha\mu}\sin{\beta}\!=\!0\label{mod}
\end{eqnarray}
specifying both derivatives of both degrees of freedom, and as such being as many as the original Dirac equations \cite{Fabbri:2024lyu}.

In polar formalism, the Dirac equations in polar form (\ref{chan}-\ref{mod}) can be re-arranged as
\begin{eqnarray}
&\nabla_{\mu}U^{\mu}\label{M2-1}\!=\!0\\
&P^{\nu}\!=\!m\cos{\beta}u^{\nu}\!-\!\frac{1}{2}\varepsilon^{\nu\mu}\nabla_{\mu}\beta\label{M2-2}\\
&\frac{1}{2}\varepsilon^{\mu\nu}\nabla_{\mu}U_{\nu}\!=\!-2m\phi^{2}\sin{\beta}\label{M2-3}
\end{eqnarray}
as proven in appendix \ref{app2}. The first is still the continuity equation for the velocity density. The second spells that the momentum can be written explicitly in terms of the chiral angle and its derivatives, which is therefore interpreted as a second type of relativistic quantum potential. It is a relativistic quantum potential in the sense that it involves only first-order derivatives of a degree of freedom. But even more, this quantum potential is relativistic also because that degree of freedom is the chiral angle, and chirality is an intrinsically relativistic character. Then, equation (\ref{M2-2}), tying momentum to quantum potential, is the relativistic Hamilton-Jacobi equation. And it is still the guidance equation.

Since the full system (\ref{M2-1}-\ref{M2-2}-\ref{M2-3}) is equivalent to (\ref{chan}-\ref{mod}), then the system of equations (\ref{M2-1}-\ref{M2-2}-\ref{M2-3}) is in its Madelung formulation. The fact that the Madelung system is made of the continuity equation and the Hamilton-Jacobi/guidance equation is no longer true since now another equation has appeared. It is the curl of the velocity density, and whereas it is never encountered in non-relativistic cases, because as shown in section \ref{1+2} it does not exist in the $3$-dimensional space, it does appear in $2$-dimensional spaces \cite{Hatifi:2024pks}. The main feature, however, is that it is formally classical, and so it can be included in the Madelung system. We can now tackle the $(1\!+\!3)$-dimensional spacetime.
\section{(1+3)-dimensional spacetime: Madelung Structure of the Dirac Equation}\label{Full}
In section \ref{1+2}, we have seen that the Dirac equations in polar form can be re-configured into a system of equations in Madelung form, containing a continuity equations and a Hamilton-Jacobi/guidance equation giving the momentum in terms of the product of mass times velocity plus contributions of the type $\nabla_{k}\ln{\phi^{2}}$ which was recognized to be the relativistic quantum potential. In section \ref{1+1}, we have seen that also contributions of the type $\nabla_{k}\beta$ might be reckoned as relativistic quantum potentials, and that another equation, the curl of the velocity density, has appeared.

These ingredients are enough, also in $(1\!+\!3)$-dimensional spacetimes. In fact, we shall see that also for the physical spacetime, the Dirac equations in polar form can be re-shaped into a system of equations consisting of the divergence and the curl of the velocity density plus a Hamilton-Jacobi/guidance equation giving the momentum in terms of $mu_{i}$ plus relativistic quantum potentials displaying first-order derivatives of the two degrees of freedom.

In the physical $(1\!+\!3)$-dimensional spacetime, Clifford gamma matrices and sigma matrices will be defined as usual and with $2i\boldsymbol{\sigma}_{ab}\!=\!\varepsilon_{abcd}\boldsymbol{\pi}\boldsymbol{\sigma}^{cd}$ defining the matrix $\boldsymbol{\pi}$ (which is the parity-odd matrix, as was denoted in section \ref{1+1}). The independent spinor bi-linears are $U^{a}$ as velocity density vector now accompanied by $S^{a}$ as spin density axial-vector, with $\Theta$ as density pseudo-scalar and $\Phi$ as density scalar, where one more time we follow all the definitions reported in reference \cite{Fabbri:2024lyu}. In polar form we have $\Theta\!=\!2\phi^{2}\sin{\beta}$ and $\Phi\!=\!2\phi^{2}\cos{\beta}$ with $U^{a}\!=\!2\phi^{2}u^{a}$ and $S^{a}\!=\!2\phi^{2}s^{a}$ in terms of module $\phi$ and chiral angle $\beta$ and with velocity $u^{a}$ and spin $s^{a}$ verifying $u_{a}u^{a}\!=\!-s_{a}s^{a}\!=\!1$ and $u_{a}s^{a}\!=\!0$ as conditions of ortho-normalization. The module and chiral angle are the two degrees of freedom of the spinorial system.

The spacetime and gauge tensorial connections, as well as the covariant derivative of the spinor field in polar form, are defined as usual \cite{Fabbri:2024lyu}. However, now we have
\begin{eqnarray}
&\nabla_{\mu}u_{\nu}\!=\!u^{\alpha}R_{\alpha\nu\mu}\ \ \ \ \ \ \ \
\ \ \ \ \ \ \ \ \nabla_{\mu}s_{\nu}\!=\!s^{\alpha}R_{\alpha\nu\mu}\label{ds-du}
\end{eqnarray}
as general identities involving both velocity and spin.

The Dirac equation is equivalent to the pair
\begin{eqnarray}
&\nabla_{\alpha}\beta\!+\!B_{\alpha}
\!-\!2P^{\nu}u_{[\nu}s_{\alpha]}\!+\!2ms_{\alpha}\cos{\beta}\!=\!0
\label{dp1}\\
&\nabla_{\alpha}\ln{\phi^{2}}\!+\!R_{\alpha}
\!-\!2P^{\rho}u^{\nu}s^{\sigma}\varepsilon_{\alpha\rho\nu\sigma}\!+\!2ms_{\alpha}\sin{\beta}\!=\!0
\label{dp2}
\end{eqnarray}
specifying all derivatives of both degrees of freedom in terms of $R_{\mu}\!=\!R_{\mu\nu}^{\phantom{\mu\nu}\nu}$ and $B_{\mu}\!=\!\frac{1}{2}\varepsilon_{\mu\alpha\nu\iota}R^{\alpha\nu\iota}$ as introduced in \cite{Fabbri:2024lyu}.

In polar formalism, the Dirac equations in polar form (\ref{dp1}-\ref{dp2}) can be re-arranged as
\begin{eqnarray}
&\nabla_{\mu}U^{\mu}\!=\!0\label{M1}\\
&\nabla^{[\alpha}U^{\nu]}
\!+\!R^{[\alpha}U^{\nu]}\!-\!R^{\alpha\nu\mu}U_{\mu}
\!+\!\varepsilon^{\alpha\nu\mu\rho}\nabla_{\mu}\beta U_{\rho}
\!+\!2\varepsilon^{\alpha\nu\mu\rho}P_{\mu}S_{\rho}\!-\!2mM^{\alpha\nu}\!=\!0\label{M2}\\
&P^{\mu}\!=\!m\cos{\beta}u^{\mu}
\!+\!\frac{1}{2}(\nabla_{\nu}\beta\!+\!B_{\nu})u^{[\nu}s^{\mu]}
\!+\!\frac{1}{2}(\nabla_{\nu}\ln{\phi^{2}}\!+\!R_{\nu})u_{\alpha}s_{\sigma}\varepsilon^{\nu\alpha\sigma\mu}\label{M3}
\end{eqnarray}
as demonstrated in appendix \ref{app3}. Equations (\ref{M1}-\ref{M2}) are the divergence and curl of the velocity density, while equation (\ref{M3}) is the Hamilton-Jacobi/guidance equation. And therefore, this system of equations is in Madelung form.
\section{Some General Comment}
We have now accomplished our goal of having the Dirac equations in polar form re-configured into Madelung form as condensed in equations (\ref{M1}-\ref{M2}-\ref{M3}). Equation (\ref{M1}) contains the divergence of the velocity density, and so up to the multiplication by $m$, it is the continuity equation, establishing the conservation law of the mass. Equation (\ref{M2}) links the curl of the velocity density, that is the vorticity, with the tensor $M^{\alpha\nu}$, which is the spinor bi-linear encoding the information about the angular momentum of the matter distribution. Equation (\ref{M3}) is the Hamilton-Jacobi/guidance equation, since it is at the same time a guidance equation, tying the momentum to the product of mass times velocity, and a Hamilton-Jacobi equation, tying the momentum to the two relativistic quantum potentials $\nabla_{\nu}\beta$ and $\nabla_{\nu}\ln{\phi^{2}}$: these two relativistic quantum potentials are quantum potentials in the sense that they are derivatives of the degrees of freedom, and they are relativistic in the sense that such derivatives are first-order; the one involving the chiral angle is relativistic also in the fact that the chiral angle itself is constructed to describe a genuinely relativistic quantity.

The quantum potentials $\nabla_{\nu}\beta$ and $\nabla_{\nu}\ln{\phi^{2}}$ are the relativistic counter-part of $Q$ and so, in the same way in which $Q$ was used to infer the trajectories of quantum particles in the de Broglie-Bohm mechanics \cite{b1, t1, b2, t2}, the two potentials $\nabla_{\nu}\beta$ and $\nabla_{\nu}\ln{\phi^{2}}$ can be used for the same purpose in the relativistic extension of Bohmian mechanics \cite{Hatifi:2024pks}. Further extensions of the results of \cite{Hatifi:2024pks} to $(1\!+\!3)$-dimensional cases are possible, although out of the scope of the present work.

That Hamilton-Jacobi and guidance equation be encoded within the same equation is possible in relativistic situations because in these cases the Hamilton-Jacobi equation is a linear dispersion relation, like the guidance equation.

The system (\ref{M1}-\ref{M2}-\ref{M3}) is physically relevant because, apart from the information contained in the quantum potentials, all equations are formally classical. No other re-configuration of the Dirac equation has this property \cite{Fabbri:2023yhl}.

As already commented, there is in general no equivalence between Hamilton-Jacobi equation and Newton equation in curved spacetimes, because the Newton equation is obtained from the Hamilton-Jacobi equation after one derivation, a passage that cannot be reversed unless specific integrability conditions are granted, and this is in general not true if the spacetime has a curvature. However, it is still possible to deduce the Newton equation. Computations necessitate the energy density tensor of the spinor, with its conservation law, as we are going to discuss in the following.
\section{Conservation Laws}
Having obtained the Madelung system (\ref{M1}-\ref{M2}-\ref{M3}) in the relativistic case with spin, in which (\ref{M3}) has the role of the Hamilton-Jacobi equation, it may be interesting now to see how we can obtain the Newton law. And in order for that to occur, we must give conserved quantities, and the corresponding conservation laws.

The three conserved quantities for the spinor field are the electric current vector
\begin{eqnarray}
&J^{\mu}\!=\!q\overline{\psi}\boldsymbol{\gamma}^{\mu}\psi\label{J}
\end{eqnarray}
the spin density tensor
\begin{eqnarray}
&S^{\alpha\nu\sigma}\!=\!\frac{i}{4}\overline{\psi}
\{\boldsymbol{\gamma}^{\alpha},\boldsymbol{\sigma}^{\nu\sigma}\}\psi\label{Se}
\end{eqnarray}
and the energy density tensor
\begin{eqnarray}
&T^{\rho\sigma}\!=\!\frac{1}{4}F^{2}g^{\rho\sigma}
\!-\!F^{\rho\alpha}\!F^{\sigma}_{\phantom{\sigma}\alpha}
\!+\!\frac{i}{2}(\overline{\psi}\boldsymbol{\gamma}^{\rho}\boldsymbol{\nabla}^{\sigma}\psi
\!-\!\boldsymbol{\nabla}^{\sigma}\overline{\psi}\boldsymbol{\gamma}^{\rho}\psi)\label{T}
\end{eqnarray}
which contains also the electrodynamic contribution. The conservation laws for these quantities are
\begin{eqnarray}
&\nabla_{\rho}J^{\rho}\!=\!0\label{Jcon}
\end{eqnarray}
together with
\begin{eqnarray}
&\nabla_{\rho}S^{\rho\mu\nu}\!+\!\frac{1}{2}T^{[\mu\nu]}\!=\!0\label{Secon}
\end{eqnarray}
and
\begin{eqnarray}
&\nabla_{\mu}T^{\mu\nu}\!=\!0\label{Tcon}
\end{eqnarray}
and they are indeed verified when the Dirac equation is valid.

In fact, when written with polar variables, the electric current vector (\ref{J}) is
\begin{eqnarray}
&J^{\mu}\!=\!qU^{\mu}
\end{eqnarray}
so that its conservation law (\ref{Jcon}) is just
\begin{eqnarray}
&\nabla_{\rho}U^{\rho}\!=\!0
\end{eqnarray}
which is the continuity equation (\ref{M1}) above. Hence, the conservation law of the electric charge and the conservation law of the mass have identical essence. The spin density tensor (\ref{Se}) is completely antisymmetric and therefore it can be written as the Hodge dual of the spin density axial-vector
\begin{eqnarray}
&S^{\alpha\nu\sigma}\!=\!\frac{1}{4}\varepsilon^{\alpha\nu\sigma\mu}S_{\mu}
\end{eqnarray}
whose conservation law (\ref{Secon}) becomes
\begin{eqnarray}
&\nabla_{\alpha}S_{\nu}\varepsilon^{\alpha\nu\sigma\mu}
\!+\!\frac{1}{2}R_{\eta\pi}^{\phantom{\eta\pi}\sigma}S_{\kappa}\varepsilon^{\eta\pi\mu\kappa}
\!-\!\frac{1}{2}R_{\eta\pi}^{\phantom{\eta\pi}\mu}S_{\kappa}\varepsilon^{\eta\pi\sigma\kappa}
\!-\!\nabla^{[\sigma}\beta S^{\mu]}
\!-\!2P^{[\sigma}U^{\mu]}\!=\!0:
\end{eqnarray}
its Hodge dual is $F^{[\alpha}s^{\nu]}\!+\!\varepsilon^{\alpha\nu\mu\rho}E_{\mu}s_{\rho}\!+\!\varepsilon^{\alpha\nu\mu\rho}P_{\mu}u_{\rho}\!=\!0$ in which the notation (\ref{E}-\ref{F}) has been used. By tracing it with $u_{\alpha}s_{\nu}$ we get $F^{\alpha}u_{\alpha}\!=\!0$ which has already been recognized as (\ref{q1}), that is the continuity equation (\ref{M1}). Therefore, the conservation of the electric current vector is implied by the conservation of the spin density tensor. As for the energy density tensor given according to
\begin{eqnarray}
&T^{\rho\sigma}\!=\!\frac{1}{4}F^{2}g^{\rho\sigma}
\!-\!F^{\rho\alpha}\!F^{\sigma}_{\phantom{\sigma}\alpha}
\!+\!P^{\sigma}U^{\rho}
\!+\!\nabla^{\sigma}\beta S^{\rho}/2
\!-\!\frac{1}{4}R_{\alpha\nu}^{\phantom{\alpha\nu}\sigma}S_{\kappa}\varepsilon^{\rho\alpha\nu\kappa}
\end{eqnarray}
its conservation law results into
\begin{eqnarray}
&U^{\rho}\nabla_{\rho}P^{\sigma}\!=\!qF^{\sigma\alpha}U_{\alpha}
\!-\!\frac{1}{2}\nabla_{\rho}\left(\nabla^{\sigma}\beta S^{\rho}
\!-\!\frac{1}{2}R_{\alpha\nu}^{\phantom{\alpha\nu}\sigma}S_{\kappa}\varepsilon^{\rho\alpha\nu\kappa}\right)\label{NS}
\end{eqnarray}
in which the conservation law of the electric current vector and of the spin density tensor (as well as the electrodynamic field equations) were used. Equation (\ref{NS}) is the Navier-Stokes equation \cite{Fabbri:2024avj}. When $\beta\!=\!0$ and $R_{\alpha\nu\mu}\!=\!0$ we can replace the velocity density with the velocity getting $u^{\rho}\nabla_{\rho}P^{\sigma}\!=\!qF^{\sigma\alpha}u_{\alpha}$ which is the Newton law with Lorentz force.
\section{Second-Order Equations}
Let us next consider the Dirac equations in polar form (\ref{dp1}-\ref{dp2}), and define the potentials
\begin{eqnarray}
&B_{\alpha}\!-\!2P^{\nu}u_{[\nu}s_{\alpha]}\!=\!M_{\alpha}\\
&R_{\alpha}\!-\!2P^{\rho}u^{\nu}s^{\sigma}\varepsilon_{\alpha\rho\nu\sigma}\!=\!\Sigma_{\alpha}
\end{eqnarray}
in terms of which (\ref{dp1}-\ref{dp2}) are re-written as
\begin{eqnarray}
&\nabla_{\alpha}\beta\!+\!M_{\alpha}\!+\!2ms_{\alpha}\cos{\beta}\!=\!0\\
&\nabla_{\alpha}\ln{\phi^{2}}\!+\!\Sigma_{\alpha}\!+\!2ms_{\alpha}\sin{\beta}\!=\!0:
\end{eqnarray}
using these two forms, we can compute the second-order derivative of the module to be
\begin{eqnarray}
&\frac{1}{4}\nabla_{\alpha}\beta\nabla^{\alpha}\beta\!-\!m^{2}\!-\!\phi^{-1}\nabla^{\alpha}\nabla_{\alpha}\phi
\!+\!\frac{1}{2}(-\nabla_{\alpha}\Sigma^{\alpha}\!+\!\frac{1}{2}\Sigma_{\alpha}\Sigma^{\alpha}
\!-\!\frac{1}{2}M_{\alpha}M^{\alpha})\!=\!0\label{II}
\end{eqnarray}
as a constraint between quantum potentials and the vectors $\Sigma_{\alpha}$ and $M_{\alpha}$, which contain the momentum. Hence, this is also a Hamilton-Jacobi equation, although one in which we have moved to a higher-order differential, with a consequent loss of information. However, having a second-order derivative equation allows one to obtain the non-relativistic limit, as we are going to show. In fact, in the case $R_{\alpha\nu\mu}\!=\!0$ and $\beta\!=\!0$ equation (\ref{II}) reduces to
\begin{eqnarray}
&-m^{2}\!-\!\phi^{-1}\nabla^{\alpha}\nabla_{\alpha}\phi\!+\!P^{i}P_{i}
\!-\!\frac{1}{2}qF^{\alpha\rho}u^{\nu}s^{\sigma}\varepsilon_{\alpha\rho\nu\sigma}\!=\!0:
\end{eqnarray}
with no time dependence, in non-relativistic limit $u_{i}\!\rightarrow\!(1,-\vec{v})$, and expanding $P^{i}P_{i}\!-\!m^{2}\!\approx\!2mH\!-\!\vec{P}\!\cdot\!\vec{P}$, we get
\begin{eqnarray}
H\!=\!\frac{1}{2m}\vec{P}\!\cdot\!\vec{P}\!+\!(V\!+\!Q)
\end{eqnarray}
in which $V\!=\!qm^{-1}\vec{B}\!\cdot\!\vec{s}/2$ with the magnetic field introduced per standard notation. This expression is equation (\ref{enerpolar}).
\section{Multi-Valuedness}
As pointed out in section \ref{non-Rel}, in order for the fundamental equations of non-relativistic quantum mechanics to have the Madelung structure it is necessary to accompany the Schr\"{o}dinger equation with the guidance equation, or more generally with a condition stating that the velocity must be the gradient of a scalar potential. In the original works of Bohm this fact is already present, although it is only with the subsequent works of Takabayasi that such a situation is completely clarified. In more recent literature, such a circumstance is known as Wallstrom objection. The Wallstrom objection retains that the requirement for which the velocity be the gradient of a scalar, that is the phase of the wave function, is not compatible with the multi-valuedness of this scalar, as is for the wave function \cite{Wallstrom:1994fp}. This is true, but only in the non-relativistic case. In fact, it is only in the non-relativistic case that, lacking a definition of velocity, the guidance equation must be postulated. In section \ref{Full}, we have seen that, in the relativistic case, there exists a natural definition of velocity. Then, the guidance equation (\ref{M3}) is derived as a consequence of the Dirac equation.

In view of the Wallstrom argument, the objection is resolved by the fact that, in relativistic cases, the spinor field is naturally multi-valued. In fact, when the spinor field is written in polar form, it is given as
\begin{eqnarray}
&\psi\!=\!\phi\ e^{-i\beta\boldsymbol{\pi}/2}
\ \boldsymbol{L}^{-1}\left(\begin{tabular}{c}
$1$\\
$0$\\
$1$\\
$0$
\end{tabular}\right)
\label{spinor}
\end{eqnarray}
for some $\boldsymbol{L}$ with the structure of a spinor transformation \cite{jl1, jl2}. This is comprehensible if we think that a relativistic spinor field has $8$ real components, and that the Dirac equation also has $8$ real components: as clear from (\ref{dp1}-\ref{dp2}), the $8$ independent equations determine all spacetime derivatives of the $2$ degrees of freedom. So, there must be $8\!-\!2\!=\!6$ components of the spinor that can not be determined by field equations. These are the velocity and spin, constrained by (\ref{ds-du}), but undetermined otherwise. Physically, this situation can be understood by thinking that while the Dirac equation must determine the general behaviour of the relativistic spinor field, there would always remain characters like the overall motion and the spin orientation that no equation can possibly fix. Intuitively, these features are related to the choices of the observer, on which the dynamical equations have no control. Such under-determination for the Dirac equation is reflected onto the under-determination of velocity and spin, and then onto $\boldsymbol{L}$ itself.

The parameters of the $\boldsymbol{L}$ matrix are the Goldstone fields of the spinor, playing the role of hidden variables. In their not being determined by field equations, they are naturally contextual \cite{Bell, CHSH, KS}.
\section{Conclusion}
In this paper, we have considered the Dirac equations written in polar form and we have expressed them in Madelung form (\ref{M1}-\ref{M2}-\ref{M3}): we observed that, after the continuity equation, involving the divergence of the velocity density, there appeared also another equation, involving the curl of the velocity density (this is a feature also of $2$-dimensional spaces, although not of $3$-dimensional spaces); we discussed how the guidance equation is also the Hamilton-Jacobi equation (a property peculiar of relativistic circumstances); we commented about the two relativistic quantum potentials (the derivatives of the degrees of freedom). The two relativistic quantum potentials are analogous to $Q$ in de Broglie-Bohm mechanics, and in analogy with \cite{b1, t1, b2, t2} particle trajectories have already been discussed in $(1\!+\!1)$-dimensions \cite{Hatifi:2024pks}: it is our hope that by means of the present formulation, all the results of \cite{Hatifi:2024pks} be soon extended to the physical spacetime.

Apart from the quantum potentials, the system of equations (\ref{M1}-\ref{M2}-\ref{M3}) is entirely classical: this is the reason why the Madelung form is fundamental to interpret quantum mechanics as a special type of classical mechanics, and again we hope that the present results will be of some help in finding our way toward such an interpretation \cite{Reddiger:2015vsa, RP}.

The Madelung system is also important for applications to complex systems \cite{W} and quantum transformations \cite{SLCVSP}.

Avenues of generalization may involve an investigation on the role of curved spaces or adding torsion \cite{Guedes:2022mdy, Pereira:2001xf}.
\appendix
\section{Equivalence of Dirac and Madelung forms: 3-dimensions}\label{app1}
The equivalence of Dirac equation and Madelung equations can be proven by showing that the Dirac equations in polar form (\ref{constraint}-\ref{trueequation}) imply and are implied by the Madelung equations (\ref{M3-1}-\ref{M3-2}). So let us suppose that (\ref{constraint}-\ref{trueequation}) be valid, and take (\ref{trueequation}) contracted with $u^{k}$: this results in
\begin{eqnarray}
&\nabla_{k}u^{k}\!+\!u^{k}\nabla_{k}\ln{\phi^{2}}\!=\!0\label{a1}
\end{eqnarray}
in which (\ref{3du}) was used. By definition of velocity, the above is just (\ref{M3-1}). Instead the Hodge dual of (\ref{trueequation}) can be worked out (recalling that $\varepsilon^{ijk}\varepsilon_{abk}\!=\!\delta^{i}_{a}\delta^{j}_{b}\!-\!\delta^{i}_{b}\delta^{j}_{a}$) to be
\begin{eqnarray}
&\varepsilon^{ijk}R_{k}\!+\!2P^{i}u^{j}\!-\!2P^{j}u^{i}
\!+\!\varepsilon^{ijk}\nabla_{k}\ln{\phi^{2}}\!=\!0:
\end{eqnarray}
taking its contraction with $u_{j}$ and using (\ref{constraint}) we obtain (\ref{M3-2}). By converse, let us consider (\ref{M3-1}-\ref{M3-2}) to be true, and take (\ref{M3-2}) contracted to $u_{i}$: the result is
\begin{eqnarray}
&P^{i}u_{i}\!=\!m\!-\!\frac{1}{4}R_{abc}\varepsilon^{abc}
\end{eqnarray}
which is (\ref{constraint}). Instead, taking (\ref{M3-2}) contracted to $2\varepsilon_{kij}u^{j}$ gives
\begin{eqnarray}
&2\varepsilon_{kij}P^{i}u^{j}\!=\!u_{k}u^{b}(\nabla_{b}\ln{\phi^{2}}\!+\!R_{b})
\!-\!(\nabla_{k}\ln{\phi^{2}}\!+\!R_{k}):
\end{eqnarray}
the above reduces to (\ref{trueequation}) after employing (\ref{M3-1}) in the form given by (\ref{a1}). Therefore, (\ref{constraint}-\ref{trueequation}) and (\ref{M3-1}-\ref{M3-2}) are equivalent.
\section{Equivalence of Dirac and Madelung forms: 2-dimensions}\label{app2}
To prove that the Dirac equations in polar form (\ref{chan}-\ref{mod}) imply and are implied by the Madelung equations (\ref{M2-1}-\ref{M2-2}-\ref{M2-3}), we start by assuming (\ref{chan}-\ref{mod}) to be valid, and take (\ref{mod}) contracted with $u^{k}$, getting
\begin{eqnarray}
&u^{k}(\nabla_{k}\ln{\phi^{2}}\!+\!R_{k})\!=\!0\label{b1}:
\end{eqnarray}
after using (\ref{2du}) and the definition of velocity, we can see that this is (\ref{M2-1}). The Hodge dual of (\ref{chan}) can be manipulated further (recalling that in this signature $\varepsilon_{ak}\varepsilon^{ik}\!=\!-\delta^{i}_{a}$) to yield
\begin{eqnarray}
&P^{a}\!=\!m\cos{\beta}u^{a}\!-\!\frac{1}{2}\varepsilon^{ak}\nabla_{k}\beta
\end{eqnarray}
which is (\ref{M2-2}). Instead, the Hodge dual of (\ref{mod}) gives
\begin{eqnarray}
&\varepsilon^{ak}(\nabla_{k}\ln{\phi^{2}}\!+\!R_{k})\!=\!2mu^{a}\sin{\beta}
\end{eqnarray}
which can be contracted with $u_{a}$ to furnish
\begin{eqnarray}
&\varepsilon^{ak}u_{a}(\nabla_{k}\ln{\phi^{2}}\!+\!R_{k})\!=\!2m\sin{\beta}\label{b2}
\end{eqnarray}
which needs more work. To this purpose, it is necessary to notice that in the $2$-dimensional spacetime, the spacetime tensorial connection can always be expressed as
\begin{eqnarray}
&R_{aik}\!=\!R_{a}g_{ik}\!-\!R_{i}g_{ak}
\end{eqnarray}
in terms of its trace. Using this fact, again (\ref{2du}) and the definition of velocity, we get
\begin{eqnarray}
&\nabla_{k}U_{i}\!=\!\nabla_{k}(2\phi^{2}u_{i})
\!=\!2\nabla_{k}\phi^{2}u_{i}\!+\!2\phi^{2}u^{a}R_{aik}
\!=\!2\nabla_{k}\phi^{2}u_{i}\!+\!2\phi^{2}u^{a}(R_{a}g_{ik}\!-\!R_{i}g_{ak})
\end{eqnarray}
and so
\begin{eqnarray}
&\frac{1}{2}\varepsilon^{ik}\nabla_{k}U_{i}\!=\!\varepsilon^{ik}u_{i}\nabla_{k}\phi^{2}
\!+\!\phi^{2}\varepsilon^{ai}u_{a}R_{i}\label{b3}
\end{eqnarray}
and now we have all we need: combining (\ref{b2}) and (\ref{b3}) allows us to arrive at
\begin{eqnarray}
&\frac{1}{2}\varepsilon^{ik}\nabla_{k}U_{i}\!=\!2m\phi^{2}\sin{\beta}
\end{eqnarray}
which is (\ref{M2-3}). Conversely, from (\ref{M2-2}), taking the Hodge dual, we immediately get (\ref{chan}). Instead, from (\ref{M2-3}), employing (\ref{b3}) and multiplying by $\varepsilon_{ab}u^{b}$ (recalling that $\varepsilon_{ab}\varepsilon^{ij}\!=\!-\delta^{i}_{a}\delta^{j}_{b}\!+\!\delta^{i}_{b}\delta^{j}_{a}$) takes us to
\begin{eqnarray}
&(\nabla_{a}\ln{\phi^{2}}\!+\!R_{a})\!-\!u_{a}u^{i}(\nabla_{i}\ln{\phi^{2}}\!+\!R_{i})
\!=\!2m\varepsilon_{ab}u^{b}\sin{\beta}:
\end{eqnarray}
this is (\ref{mod}) when (\ref{M2-1}) is used in the form given by (\ref{b1}). Therefore, equations (\ref{chan}-\ref{mod}) and (\ref{M2-1}-\ref{M2-2}-\ref{M2-3}) are equivalent.
\section{Equivalence of Dirac and Madelung forms: 4-dimensions}\label{app3}
The proof of the equivalence of the Dirac equations in polar form (\ref{dp1}-\ref{dp2}) and the Madelung equations (\ref{M1}-\ref{M2}-\ref{M3}) is considerably more complex, so we will start by recalling a few definitions: first of all we recall from \cite{Fabbri:2024lyu} that
\begin{eqnarray}
&M_{ab}\!=\!2\phi^{2}(\cos{\beta}u^{j}s^{k}\varepsilon_{jkab}\!+\!\sin{\beta}u_{[a}s_{b]})\label{M}
\end{eqnarray}
so that, by means of the definitions of velocity and spin, we can write (\ref{M1}-\ref{M2}-\ref{M3}) according to
\begin{eqnarray}
&u^{\mu}(\nabla_{\mu}\ln{\phi^{2}}\!+\!R_{\mu})\!=\!0\label{m1}\\
&(\nabla^{[\alpha}\ln{\phi^{2}}\!+\!R^{[\alpha})u^{\nu]}
\!+\!\varepsilon^{\alpha\nu\mu\rho}(\nabla_{\mu}\beta\!+\!B_{\mu})u_{\rho}
\!+\!2\varepsilon^{\alpha\nu\mu\rho}P_{\mu}s_{\rho}\!-\!2m(\cos{\beta}u_{j}s_{k}\varepsilon^{jk\alpha\nu}\!+\!\sin{\beta}u^{[\alpha}s^{\nu]})\!=\!0\label{m2}\\
&P^{\mu}\!=\!m\cos{\beta}u^{\mu}
\!+\!\frac{1}{2}(\nabla_{\nu}\beta\!+\!B_{\nu})u^{[\nu}s^{\mu]}
\!+\!\frac{1}{2}(\nabla_{\nu}\ln{\phi^{2}}\!+\!R_{\nu})u_{\alpha}s_{\sigma}\varepsilon^{\nu\alpha\sigma\mu}\label{m3}
\end{eqnarray}
where again (\ref{ds-du}) were used. Because this passage involves only definitions, the system (\ref{M1}-\ref{M2}-\ref{M3}) is equivalent to the system (\ref{m1}-\ref{m2}-\ref{m3}), and hence the equivalence of (\ref{dp1}-\ref{dp2}) and (\ref{M1}-\ref{M2}-\ref{M3}) is reduced to prove the equivalence of (\ref{dp1}-\ref{dp2}) and (\ref{m1}-\ref{m2}-\ref{m3}). Additionally, we can introduce the auxiliary vectors
\begin{eqnarray}
&2E_{\mu}\!=\!B_{\mu}\!+\!\nabla_{\mu}\beta\!+\!2ms_{\mu}\cos{\beta}\label{E}\\
&2F_{\mu}\!=\!R_{\mu}\!+\!\nabla_{\mu}\ln{\phi^{2}}\!+\!2ms_{\mu}\sin{\beta}\label{F}:
\end{eqnarray}
with them, the Dirac equations in polar form (\ref{dp1}) and (\ref{dp2}) become
\begin{eqnarray}
&E_{\mu}\!=\!P^{\nu}u_{[\nu}s_{\mu]}\label{d1}\\
&F_{\mu}\!=\!P^{\rho}u^{\nu}s^{\sigma}\varepsilon_{\mu\rho\nu\sigma}\label{d2}
\end{eqnarray}
while the Madelung equations (\ref{m1}), the Hodge dual of (\ref{m2}) and (\ref{m3}) become
\begin{eqnarray}
&u^{\mu}F_{\mu}\!=\!0\label{q1}\\
&\varepsilon_{\alpha\nu\sigma\pi}F^{\alpha}u^{\nu}\!-\!E_{[\sigma}u_{\pi]}
\!-\!P_{[\sigma}s_{\pi]}\!=\!0\label{q2}\\
&P^{\mu}\!=\!E_{\nu}u^{[\nu}s^{\mu]}
\!+\!F_{\nu}u_{\alpha}s_{\sigma}\varepsilon^{\nu\alpha\sigma\mu}\label{q3}
\end{eqnarray}
and the problem is reduced to prove the equivalence of (\ref{d1}-\ref{d2}) and (\ref{q1}-\ref{q2}-\ref{q3}). For this, we begin by assuming (\ref{d1}-\ref{d2}): having (\ref{d2}) contracted with $u^{\mu}$ gives
\begin{eqnarray}
&u^{\mu}F_{\mu}\!=\!P^{\rho}u^{\mu}u^{\nu}s^{\sigma}\varepsilon_{\mu\rho\nu\sigma}\!=\!0
\end{eqnarray}
which is (\ref{q1}). Having (\ref{d2}) contracted with $\varepsilon^{\mu\nu\sigma\pi}u_{\nu}$ gives
\begin{eqnarray}
&\varepsilon^{\mu\nu\sigma\pi}F_{\mu}u_{\nu}\!=\!P^{[\sigma}s^{\pi]}
\!+\!P^{\nu}u_{\nu}u^{[\pi}s^{\sigma]}\label{Fue}
\end{eqnarray}
while having (\ref{d1}) contracted with $u^{\mu}$ and $s^{\mu}$ gives
\begin{eqnarray}
&E_{\mu}u^{\mu}\!=\!-P^{\nu}s_{\nu}\label{Eu}\\
&E_{\mu}s^{\mu}\!=\!-P^{\nu}u_{\nu}\label{Es}:
\end{eqnarray}
we can now use (\ref{Fue}) and (\ref{d1}) to evaluate the left side of (\ref{q2}) as
\begin{eqnarray}
&\varepsilon^{\mu\nu\sigma\pi}F_{\mu}u_{\nu}\!-\!E^{[\sigma}u^{\pi]}
\!-\!P^{[\sigma}s^{\pi]}
\!=\!P^{[\sigma}s^{\pi]}
\!+\!P^{\nu}u_{\nu}u^{[\pi}s^{\sigma]}
\!+\!P_{\omega}u^{[\omega}s^{\pi]}u^{\sigma}
\!-\!P_{\omega}u^{[\omega}s^{\sigma]}u^{\pi}
\!-\!P^{[\sigma}s^{\pi]}\!=\!0
\end{eqnarray}
showing that (\ref{q2}) is valid. And we can use (\ref{Eu}-\ref{Es}) and (\ref{Fue}) to evaluate the right side of (\ref{q3}) and see that
\begin{eqnarray}
&E_{\nu}u^{[\nu}s^{\mu]}
\!+\!F_{\nu}u_{\alpha}s_{\sigma}\varepsilon^{\nu\alpha\sigma\mu}
\!=\!-P^{\nu}s_{\nu}s^{\mu}\!+\!P^{\nu}u_{\nu}u^{\mu}
\!+\!P^{[\sigma}s^{\mu]}s_{\sigma}
\!+\!P^{\nu}u_{\nu}u^{[\mu}s^{\sigma]}s_{\sigma}\!=\!P^{\mu}
\end{eqnarray}
showing that (\ref{q3}) is valid indeed. Conversely, let us take (\ref{q1}-\ref{q2}-\ref{q3}): having (\ref{q2}) contracted with $\varepsilon^{\sigma\pi\eta\mu}u_{\mu}$ and using (\ref{q1}) yields
\begin{eqnarray}
&F^{\eta}\!=\!\varepsilon^{\eta\sigma\mu\pi}P_{\sigma}u_{\mu}s_{\pi}
\end{eqnarray}
which is (\ref{d2}). On the other hand, having (\ref{q2}) contracted with $u^{\sigma}$ gives
\begin{eqnarray}
&E_{\sigma}u^{\sigma}u_{\pi}\!-\!E_{\pi}\!+\!P_{\sigma}u^{\sigma}s_{\pi}\!=\!0
\end{eqnarray}
while having (\ref{q3}) contracted with $s_{\mu}u^{\omega}$ gives
\begin{eqnarray}
&P^{\mu}s_{\mu}u^{\omega}\!=\!-E_{\nu}u^{\nu}u^{\omega}
\end{eqnarray}
and these last two equations can be combined to provide
\begin{eqnarray}
&E_{\pi}\!=\!E_{\sigma}u^{\sigma}u_{\pi}\!+\!P_{\sigma}u^{\sigma}s_{\pi}
\!=\!P^{\sigma}(u_{\sigma}s_{\pi}\!-\!s_{\sigma}u_{\pi})
\end{eqnarray}
which is (\ref{d1}). In conclusion, we have demonstrated that the two systems (\ref{d1}-\ref{d2}) and (\ref{q1}-\ref{q2}-\ref{q3}) are equivalent.

We notice that, differently from the lower-dimensional cases, in the present case the Madelung system (\ref{q1}-\ref{q2}-\ref{q3}) does not consist of the exact same number of equations contained in the Dirac system (\ref{d1}-\ref{d2}), with $11$ equations in the former and $8$ equations in the latter: this apparent contradiction can be solved showing that $3$ equations are in fact repeated twice. This can be seen by taking advantage of the fact that there always exists a frame in which $u^{0}\!=\!1$ and $s^{3}\!=\!1$ \cite{Fabbri:2023yhl}: in this frame the system (\ref{q1}-\ref{q2}-\ref{q3}) is explicitly given by
\begin{eqnarray}
&F^{0}\!=\!0\label{1}\\
&F^{1}\!+\!P^{2}\!=\!0\ \ \ \ \ \ \ \ F^{2}\!-\!P^{1}\!=\!0\ \ \ \ \ \ \ \ F^{3}\!=\!0\ \ \ \
\ \ \ \ E^{3}\!-\!P^{0}\!=\!0\ \ \ \ \ \ \ \ E^{2}\!=\!0\ \ \ \ \ \ \ \ E^{1}\!=\!0\label{2}\\
&P^{0}\!=\!E^{3}\ \ \ \ \ \ \ \ P^{1}\!=\!F^{2}\ \ \ \
\ \ \ \ P^{2}\!=\!-F^{1}\ \ \ \ \ \ \ \ P^{3}\!=\!E^{0}\label{3}
\end{eqnarray}
showing that equations $P^{1}\!=\!F^{2}$, $P^{2}\!=\!-F^{1}$, $P^{0}\!=\!E^{3}$ are repeated twice, once in (\ref{2}) and once in (\ref{3}), and these are the $3$ equations that amount to the redundancy. There is, however, no physical significance in this repetition.
\vspace{10pt}

\textbf{Funding}. This work is carried out in the framework of the INFN Research Project QGSKY and funded by Next Generation EU via the project ``Geometrical and Topological effects on Quantum Matter (GeTOnQuaM)''.

\

\textbf{Data availability}. The manuscript does not have associated data in any repository.

\

\textbf{Conflict of interest}. There is no conflict of interest.

\end{document}